\documentclass[conference]{IEEEtran}
\IEEEoverridecommandlockouts
\usepackage{cite}
\usepackage{amsmath,amssymb,amsfonts}
\usepackage{algorithmic}
\usepackage{amsmath}
\usepackage{graphicx}
\usepackage{textcomp}
\usepackage{xcolor}
\usepackage{booktabs}
\usepackage[colorlinks=true,bookmarks=false,citecolor=blue,urlcolor=blue]{hyperref} 
\usepackage{multirow}
\usepackage{tabularx}

\usepackage[pscoord]{eso-pic} 

\def\BibTeX{{\rm B\kern-.05em{\sc i\kern-.025em b}\kern-.08em
    T\kern-.1667em\lower.7ex\hbox{E}\kern-.125emX}}

\begin{document}

\title{Reinforcement-Learning based routing for packet-optical networks with hybrid telemetry}

\author{
\IEEEauthorblockN{A. L. Garc\'{i}a Navarro, N. Koneva\IEEEauthorrefmark{1}, A. S\'{a}nchez-Maci\'{a}n\IEEEauthorrefmark{1},\\ J. A. Hern\'{a}ndez\IEEEauthorrefmark{1}, \'{O}. Gonz\'{a}lez de Dios\IEEEauthorrefmark{2}, J. M. Rivas-Moscoso\IEEEauthorrefmark{2} 
}
\IEEEauthorblockA{\IEEEauthorrefmark{1}Dept. Ing. Telemática, Universidad Carlos III de Madrid, Spain.}
\IEEEauthorblockA{\IEEEauthorrefmark{2}Telefonica I+D, Madrid, Spain}}



\maketitle

\begin{abstract}
This article provides a methodology and open-source implementation of Reinforcement Learning algorithms for finding optimal routes in a packet-optical network scenario. The algorithm uses measurements provided by the physical layer (pre-FEC bit error rate and propagation delay) and the link layer (link load) to configure a set of latency-based rewards and penalties based on such measurements. Then, the algorithm executes Q-learning based on this set of rewards for finding the optimal routing strategies. It is further shown that the algorithm dynamically adapts to changing network conditions by re-calculating optimal policies upon either link load changes or link degradation as measured by pre-FEC BER.
\end{abstract}


\section{Introduction}

Path optimization for traffic flows is a method available to enhance the quality of experience (QoE) perceived by users. Network automation facilitates this goal by monitoring and collecting telemetry information and network states for both optical and packet-based data. Advanced artificial intelligence, machine learning, or other intelligent algorithms are then applied to evaluate network performance. Subsequently, decisions are made and actions are taken over the network to prevent or correct possible performance issues that affect perceived QoE. This process is commonly referred to as closed-loop automation. Network automation is supported by various processes, including the implementation of Software Defined Networking (SDN), aimed at moving towards a zero-touch network and service management (ZSM) approach~\cite{gallego_zsm}. While having valid and up-to-date information is important, choosing the appropriate intelligent model to detect and correct performance problems allows for making the best decisions to optimize network operation.

In this article, we focus on the second stage of zero-touch networking: optimal path decision based on both optical and packet-based telemetry information, using the well-known Reinforcement Learning (RL) methodology, which enables optimal network configurations by allowing the control plane to learn from its interactions with the network and make decisions without human intervention. In the past, RL has been proposed to enable ZTN by providing the network with the ability to learn from its own experience and make decisions without human input~\cite{ZTN}. Several good surveys in this area are available~\cite{survey1, survey2, survey3}.

In particular, we provide a methodology for generating rewards in an RL environment, where such rewards are based on both optical telemetry information (i.e. pre-FEC BER) and packet routing measurements (i.e. latency and queue occupation). Open-source code is also provided for the interested reader willing to replicate the experiments and incorporate new features into the algorithm~\cite{alex_github}.

\section{Background and methodology}


In general, Reinforcement learning (RL) is a type of AI/ML strategy in which an agent learns to behave in an environment by trial and error, that is, by making decisions and receiving positive rewards (or penalties as negative rewards). The agent is rewarded for taking actions that lead to desired outcomes and penalized when undesired outcomes occur. Over time, the agent learns to take the best actions in each situation that maximize its rewards (or minimize penalties)~\cite{rl_explanation}. RL is effective for a variety of tasks in optical networks, including resource allocation (wavelengths and bandwidth), traffic engineering of flows to minimize congestion and resiliency against fault management~\cite{rl_optical_routing,rl_provisioning,rl_gym}. 

The formulation of RL problems require to define: 
\begin{itemize}
    \item A set of states $S$ which are representations of the environment at a given point in time. 
    \item A set of Actions $A$ that can be taken by the agent at a given state $s\in S$.
    \item Rewards which are feedback signals that the agent receives from the environment after taking an action in a given state. Rewards can be positive or negative. 
    \item The policy $\pi$ which maps states to actions. 
\end{itemize}

The goal of reinforcement learning is to find a policy $\pi$ that maximizes the agent's expected discounted return in the long term. The agent can do this by trial and error. It tries different actions at different states and observes the received rewards. Over time, the agent learns to take those actions that lead to higher expected discounted returns. There are many libraries with functions that different RL algorithms already coded, both in R and Python frameworks~\cite{rl_gym}. Examples, for open-source programming language R, include \texttt{contextual}, \texttt{ReinforcementLearning} and \texttt{MDPtoolbox}.

In this work, we use the \texttt{igraph} library for building network topologies and an implementation of the \texttt{Q-learning} algorithm for finding the optimal routing policy in a packet-optical network where a Path Computation Elements (PCE) decides the best route selection for every source-destination pair, using both optical metrics (pre-FEC bit error rate) and packet latency measurements (including propagation delay and link load). The code is publicly available in Github for further developments by the research community~\cite{alex_github}.

The Q-learning algorithm, a form of model-free reinforcement learning, updates its value function based on an equation that considers the immediate reward received for an action, plus the maximum future rewards. The Q-value of a state-action pair $(s, a)$ is updated as follows:

\begin{equation}
Q(s, a) \leftarrow Q(s, a) + \alpha \left[ R(s, a) + \gamma \max_{a'} Q(s', a') - Q(s, a) \right]
\end{equation}
where:
\begin{itemize}
    \item $Q(s, a)$ denotes the current estimate of the value of action $a$ in state $s$.
    \item $\alpha$ is the learning rate, determining the impact of new information on the existing Q-value.
    \item $R(s, a)$ is the immediate reward received after taking action $a$ in state $s$.
    \item $\gamma$ is the discount factor, which balances the importance of immediate and future rewards.
    \item $\max_{a'} Q(s', a')$ represents the maximum predicted reward achievable in the next state $s'$, considering all possible actions $a'$.
\end{itemize}

The next section shows the applicability of code~\cite{alex_github} in a few network scenarios.

\section{Simulation scenario and RL-based solution}
\subsection{Example on a small network topology}
\begin{figure}[!htbp]
    \centering
    \includegraphics[width=\columnwidth]{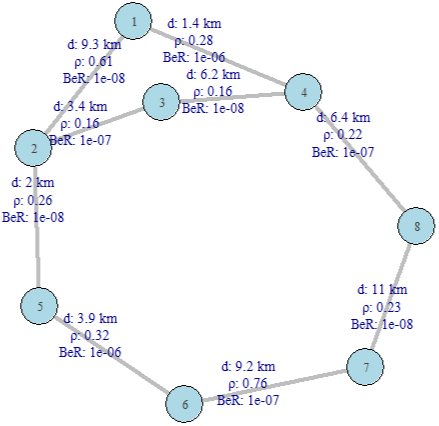}
    \caption{8-node topology example}
    \label{fig:topology8}
\end{figure}

Let us consider the network topology of Fig.~\ref{fig:topology8}, which comprises 8 nodes and 9 links. We assume that all nodes report telemetry measurements regularly to the control plane, in which our Reinforcement Learning algorithm is running to decide best routing strategies between nodes. In particular, each node reports: measured pre-FEC bit error rate (BER) and link load, denoted as $BER_i$ and $\rho_i$ for the $i$-th link. This information, together with the link distance $d_i$ in kilometer will be used by the RL algorithm to create negative rewards (penalties), as it follows: 
\begin{itemize}
    \item Propagation delay adds a penalty of $d_i \times 5~\mu s/km$, that is, the classical $5~\mu s$ signal propagation latency per kilometer of silica fiber.
    \item Traversing a given link also adds a latency penalty of  $1~\mu s \cdot \frac{1}{1-\rho_i}$, which is the average transmission and queuing delay of a 1250-byte packet transmitted over a 10~Gb/s link with load $\rho_i$ (for a classical M/M/1 queue).
    \item Monitored Pre-FEC BER adds a penalty of $1000~\mu s$ if the BER value of that link is $10^{-4}$ or above; $50~\mu s$ if the link's BER is in the range between $10^{-5} < BER_i < 10^{-4}$; or a $0~\mu s$ penalty otherwise. 
\end{itemize}

As shown, traversing each link adds both packet-based penalty, propagation-delay penalty, and optical pre-FEC BER-related penalty to either encourage or discourage links in a path. This set of rules is crafted as Reward matrices for taking action $a$ in state $s$ (i.e. $R(s,a)$ state-action pair), and inputs the Q-learning algorithm. Finally, node connectivity is also included as a bi-dimensional Matrix $P(s,s')$ which contains the probability of jumping from state (or node) $s$ to $s'$. 


\begin{table*}
\centering
\caption{Optimal path (policy) selection under normal conditions (low pre-FEC BER values) along with secondary routes after degradation of links 3-4 and 7-8}
\begin{tabular*}{\textwidth}{@{\extracolsep{\fill}}|cccccccccc|}
\hline
  \multicolumn{10}{|c|}{Dest} \\
Source & Path & 1 & 2 & 3 & 4 & 5 & 6 & 7 & 8\\ \hline
\multirow{2}{*}{1} &primary & - & 1-2 & 1-4-3 & 1-4 & 1-2-5 & 1-2-5-6 & 1-4-8-7 & 1-4-8\\
 &secondary & - & 1-2 & 1-4-3 & 1-4 & 1-2-5 & 1-2-5-6 & 1-4-8-7 & 1-4-8\\ \hline
\multirow{2}{*}{2} &primary & 2-1 & - & 2-3 & 2-1-4 & 2-5 & 2-5-6 & 2-5-6-7 & 2-1-4-8\\
 &secondary & 2-1 & - & 2-3 & 2-1-4 & 2-5 & 2-5-6 & 2-5-6-7 & 2-1-4-8\\ \hline
\multirow{2}{*}{3} &primary & 3-4-1 & 3-2 & - & 3-4 & 3-2-5 & 3-2-5-6 & 3-4-8-7 & 3-4-8\\
 &secondary & 3-2-1 & 3-2 & - & \textbf{3-2-1-4} & 3-2-5 & 3-2-5-6 & \textbf{3-2-5-6-7} & \textbf{3-2-1-4-8}\\ \hline
\multirow{2}{*}{4} &primary & 4-1 & 4-1-2 & 4-3 & - & 4-1-2-5 & 4-1-2-5-6 & 4-8-7 & 4-8\\
 &secondary & 4-1 & 4-1-2 & \textbf{4-1-2-3} & - & 4-1-2-5 & 4-1-2-5-6 & \textbf{4-1-2-5-6-7} & 4-8\\ \hline
\multirow{2}{*}{5} &primary & 5-2-1 & 5-2 & 5-2-3 & 5-2-1-4 & - & 5-6 & 5-6-7 & 5-2-1-4-8\\
 &secondary & 5-2-1 & 5-2 & 5-2-3 & 5-2-1-4 & - & 5-6 & 5-6-7 & 5-2-1-4-8\\ \hline
\multirow{2}{*}{6} &primary & 6-5-2-1 & 6-5-2 & 6-5-2-3 & 6-5-2-1-4 & 6-5 & - & 6-7 & 6-7-8\\
 &secondary & 6-5-2-1 & 6-5-2 & 6-5-2-3 & 6-5-2-1-4 & 6-5 & - & 6-7 & \textbf{6-5-2-1-4-8}\\ \hline
\multirow{2}{*}{7} &primary & 7-8-4-1 & 7-6-5-2 & 7-8-4-3 & 7-8-4 & 7-6-5 & 7-6 & - & 7-8\\
 &secondary & \textbf{7-6-5-2-1} & 7-6-5-2 & \textbf{7-6-5-2-3} & \textbf{7-6-5-2-1-4} & 7-6-5 & 7-6 & - & \textbf{7-6-5-2-1-4-8}\\ \hline
\multirow{2}{*}{8} &primary & 8-4-1 & 8-4-1-2 & 8-4-3 & 8-4 & 8-4-1-2-5 & 8-7-6 & 8-7 & -\\
 &secondary & 8-4-1 & 8-4-1-2 & \textbf{8-4-1-2-3} & 8-4 & 8-4-1-2-5 & \textbf{8-4-1-2-5-6} & \textbf{8-4-1-2-5-6-7} & -\\
\hline
\end{tabular*}
\label{fig:normal_conditions}
\end{table*}

Table~\ref{fig:normal_conditions} shows the optimal routing policy decided by our RL algorithm for the 8-node topology of Fig.~\ref{fig:topology8}. The policy finds the best next hop and primary path from source to destination taking into account the rewards for a given pre-FEC BER, propagation delay, and link load. In the example of Fig.\ref{fig:topology8}, all links operate with good quality optical links, i.e. pre-FER under $10^{-5}$, hence only propagation delay and link load contribute to finding the best primary end-to-end path. However, if the optical quality of a link degrades, then the RL algorithm finds an alternative or secondary route. This is the situation observed in the table ("secondary" rows) when links 3-4 and 7-8 experience degraded pre-FEC BER. As shown, the RL algorithm finds new routes that avoid the use of such low-quality links (marked in bold font).

\subsection{Extended example on a large topology: Tokyo MAN}

\begin{figure}[!htbp]
    \centering
    \includegraphics[width=0.9\columnwidth]{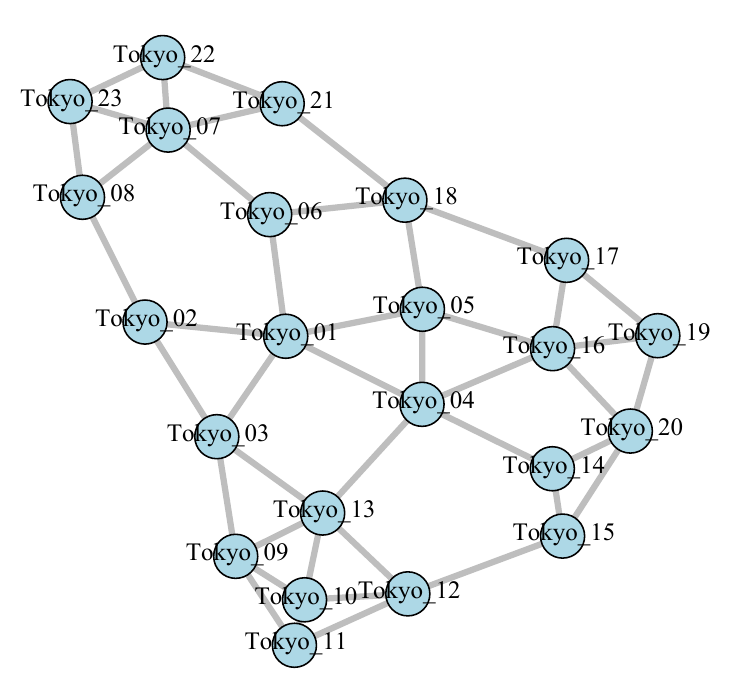}
    \caption{Tokyo topology example}
    \label{fig:tokyo_topology}
\end{figure}

Fig.~\ref{fig:tokyo_topology} shows the 23-node MAN topology for Tokyo \cite{TokyoTopology} for testing our algorithm. 

In this detailed examination, given the extensive scale and intricate nature of the network topology, our analysis will be concentrated on a select number of routing paths rather than the entire network. Key routes, including but not limited to the journey from Router 1 to Router 22, will be scrutinized. The optimal paths for these specific routes under normal conditions (primary) are depicted in Table~\ref{fig:normal_conditions_tokyo} 
that also includes secondary routes after degradation of links 1-6, 1-4, and 10-11.

\begin{table}[!htbp]
\centering
\label{tab:paths}
\caption{Optimal policy: Primary (in normal operation) and secondary (after degradation of links 1-6, 1-4, and 10-11 for the Tokyo topology).}
\begin{tabular}{|c c|c|c|c|c|}
\hline
From & To & Primary & Reward & Secondary & Reward \\ \hline
1 & 22 & 1-6-7-22 & -123.55 & 1-5-18-21-22 & -151.00 \\
4 & 7 & 4-1-6-7 & -121.88 & 4-5-18-21-7 & -157.61 \\
4 & 11 & 4-13-10-11 & -121.01 & 4-13-12-11 & -122.83 \\
1 & 19 & 1-4-16-19 & -101.81 & 1-5-16-19 & -108.21 \\
\hline
\end{tabular}
\label{fig:normal_conditions_tokyo}
\end{table}

\section{Summary and discussion}
This work has introduced an algorithm for finding optimal routing policies taking into account different physical and optical metrics (link distance and pre-FEC BER) and congestion-based ones (link load), based on Reinforcement Learning. The algorithm is open-source and can be modified to enforce finding optimal paths based on other metrics, like using high-speed links or low power consumption ones. Future work will include the application of the algorithm to other topologies that incoporate different segments of the network and also adding other QoE parameters as inputs for the algorithm.



\section*{Acknowledgment}
The authors would like to acknowledge the support of Spanish projects ITACA (PDC2022-133888-I00) and 6G-INTEGRATION-3 (TSI-063000-2021-127) and EU project SEASON (grant no. 101096120).



\footnotesize

\bibliographystyle{IEEEtran}


\begin{thebibliography}{99} 
\bibitem{gallego_zsm}
J. Gallego-Madrid, "Machine learning-based zero-touch network and service management: a survey,” Digit. Commun. Networks 8, 105–123 (2022).

\bibitem{ZTN}
O. Iacoboaiea, J. Krolikowski, Z. Ben Houidi, and D. Rossi, "From Design to Deployment of Zero-touch Deep Reinforcement Learning WLANs," arXiv:2207.06172, 2022.

\bibitem{survey1}
S. Barzegar, M. Ruiz and L. Velasco, "Autonomous Flow Routing for Near Real-Time Quality of Service Assurance," in IEEE Transactions on Network and Service Management.

\bibitem{survey2}
C. Hernández-Chulde et al, "Experimental evaluation of a latency-aware routing and spectrum assignment mechanism based on deep reinforcement learning," J. Opt. Commun. Netw.  15, 925-937 (2023).

\bibitem{survey3}
Z. Mammeri, "Reinforcement Learning Based Routing in Networks: Review and Classification of Approaches," in IEEE Access, vol. 7, pp. 55916-55950, 2019.

\bibitem{alex_github}
A. L. García Navarro, “Packet-Optical Latency-based RL,” \url{https://github.com/alexgaarciia/PacketOpticalLatencyRL} (2023).

\bibitem{rl_explanation}
M. Naeem, S. Rizvi, and A. Coronato, "A Gentle Introduction to Reinforcement Learning and its Application in Different Fields," IEEE Access, vol. 8, pp. 209320-209344, Jan. 2020.




\bibitem{rl_optical_routing} Y. Pointurier, F. Heidari, "Reinforcement learning based routing in all-optical networks," In Proc. BROADNETS 2007.

\bibitem{rl_provisioning} J. Momo Ziazet, B. Jaumard, "Deep Reinforcement Learning for Network Provisioning in Elastic Optical Networks," In Proc. ICC 2022.

\bibitem{rl_gym} C. Natalino and P. Monti, "The Optical RL-Gym: An open-source toolkit for applying reinforcement learning in optical networks," in Proc. ICTON 2020.

\bibitem{TokyoTopology} T. Tachibana et al, "Metropolitan Area Network Model Design Using Regional Railways Information for Beyond 5G Research" in IEICE Trans. on Comm., vol E106.B, no. 4, pp. 296-306, 2023.























\end{thebibliography}


\end{document}